\documentclass[a4paper,11pt]{article}
\usepackage[english]{babel}
\usepackage[latin3]{inputenc}
\usepackage[T1]{fontenc}
\usepackage{textcomp}
\usepackage{epsfig}
\usepackage{latexsym}
\usepackage{graphicx}
\usepackage{amsfonts,amssymb,bm}
\usepackage{color}
\newcommand{\D}{{\rm d}}
\newcommand{\sign}{{\rm sign\,}}
\newcommand{\artanh}{{\rm artanh\,}}

\begin{document}
\title{An extension of the principle of relativity for one-dimensional space}
\author{Josep Llosa\\
\small Departament de Física Fonamental, Universitat de Barcelona, Spain}

\maketitle

\begin{abstract}

\noindent
The class of accelerated reference frames has been studied, on teh basis of Fermi-Walker coordinates; both in the cases of uniform and arbitrary accelerations. In the first case, explicit formulae for the transformation of coordinates have been derived and, in both cases, we have also obtained the infinitesimal generators and their commutation relations. The outcome has been an extension of the Poincaré algebra (in 1+1 dimensions), which is infinite dimensional in the case of general acceleration. This extension turns out to be trivial, in the sense that it is abelian.

PACS number: 02.40.Ky, 02.20.Tw, 02.20.Sv, 04:20.Cv,  
\end{abstract}

\section{Introduction\label{S0}}
``Is it conceivable that the principle of relativity also applies to systems that are accelerated relative to each other?'' as Einstein poses in section V of ``On the relativity principle and the conclusions drawn from it'' \cite{Einstein1907}. 
And later on in the same paper he states the equivalence of a [uniform] gravitational field and the corresponding acceleration of a reference system and remarks that ``This assumption extends the principle of relativity to the uniformly accelerated translational motion of the reference system.''

In his endeavour to his relativistic theory of gravitation, that is known as the General Theory of Relativty, he will successively:
(a) formulate the equivalence principle in its weak and strong forms,
(b) establish the connexion between the gravitational potential and the [variable] speed of light, 
(c) advance the need for a generalisation of the theory of spacetime which is known as theory of relativity [in the narrow sense] \cite{Einstein1911},
(d) adopt the mighty spacetime formalism of Minkowski and introduce ten gravitational potentials (instead of only one), that are associated with the spacetime metrics \cite{Einstein1913a},\cite{Einstein1914}
(e) and embrace the principle of general covariance which, more or less oportunistically, he will weaken, drop and finally take again in his seminal 1916 paper \cite{Einstein1916}. 

Apparently the goal was the theory of relativity to abide gravitation and it seems that the idea of generalizing the theory of relativity to accelerated motions was abandoned in favour of the principle of general covariance. 
The latter obviously allows for a much larger invariance group, namely spacetime diffeomorphisms; however, as keenly pointed out and commented by Fock \cite{Fock}, such an invariance lacks of physical content. 

Nevertheless, the above quoted question posed by Einstein in the onset \cite{Einstein1907} seems legitimate. In Newtonian mechanics, the extended family of reference frames is the class of {\em rigid reference frames}, in which Newton laws hold provided that the necessary inertial force fields ---dragging, Coriolis, centrifugal, \ldots--- are also included. The transformation law connecting the coordinates of two frames in this class is
$$  x^{\prime i} = R^i_{\,j}(t)  x^j + s^i(t)\,, \qquad \qquad t^\prime = t + t_0 \,,  $$
where the functions $\, R^i_{\,j}(t)$ yield an orthogonal 3-dimensional matrix and $\,s^i(t)\,$ are arbitrary functions.

Our goal here is to set up a theory of reference frames extending the principle of relativity to accelerated relative motions. A reference frame will consist of an outfit of synchronized clocks that are at rest in the reference space. The geometry of the reference space is based on the use of stationary rigid rods that determine a metric, not necessarily Euclidean, i. e. flat.

This framework, supplemented with the requirement that the space metric is given by the infinitesimal radar distance \cite{Born}, \cite{Landau} ---that is, the two-way speed of light in vacuo is a constant--- do not allow going too far in the case of 3+1 spacetime.
Very soon \cite{Herglotz} it was clear that, even in Minkowski spacetime, the only permited rigid motions (in the sense of Born) are: (i) rectilinear uniform motions, (i) uniform rotational motions around a fixed axis and (iii) arbitrary accelerated motions without rotation.
It was clear that arbitrary rotational motions posed a genuine obstruction in two respects: the inviability of synchronizing stationary clocks along a path surrounding the rotation axis and in that the space geometry associated to the infinitesimal radar distance is neither Euclidean nor rigid \cite{Ehrenfest}.

To avoid this ``no go'' we shall here trivialize rotations and attack a simplified problem by considering a 1+1 Minkowski spacetime. 
We shall use Fermi-Walker reference frames to embody accelerated systems of reference and prove that these are the only ones which are synchronous and whose space is flat. We shall then derive the general transformation law connecting two Fermi-Walker reference frames in steadily accelerated motion. These transformations forming a 4-parameter Lie group, we shall then obtain the infinitesimal generators and the corresponding Lie algebra, which is an extension of the Poincaré algebra in 1+1 dimensions. We shall finally turn to non-uniformly accelerated relative motions, characterize them as {\em generalized isometries } \cite{Bel93} of the spacetime metric and, by solving the corresponding generalised Killing equation, we shall obtain an infinite dimensional extension of Poincaré algebra which is the mathematical counterpart of the extended relativity principle in 1+1 spacetime.

\section{Fermi-Walker coordinates  \label{S1}}
Let  $z^\mu(\tau) = (x(\tau), t(\tau))\,$ be a timelike worldline in $1+1$ Minkowski spacetime, which we shall take as the {\em space origin} (we take $c=1\,,$ $\mu=1,2\,$, $x^1= x\,, \quad x^2 =t$), $u^\mu = \dot{z}^\mu(\tau)\,$ the unit velocity vector, $a^\mu = \ddot{z}(\tau) \,$ the proper acceleration and $l^\mu$ the corresponding unit vector and $\, a = \sqrt{a^\mu a_\mu}\, $ is the scalar acceleration. 

The unit vectors $u^\mu$ and $l^\mu$ are an orthonormal basis in $1+1$ Minkowski spacetime and a function $\zeta(\tau)$ exists such that
\begin{equation}  \label{E1}
u^\mu = (\sinh \zeta, \cosh \zeta ) \,, \qquad    l^\mu = (\cosh \zeta, \sinh \zeta ) \,, \qquad a = \frac{\D\zeta}{\D \tau}
\end{equation}
( $\zeta$ is sometimes called {\em rapidity} \cite{Rindler}).
These two vectors constitute the intrinsic base associated to the curve $z^\mu(\tau)$ and the scalar acceleration $a(\tau)$ is its extrinsic curvature:
\begin{equation}  \label{E2}
\frac{\D u^\mu}{\D \tau} = a\,l^\mu  \,, \qquad \qquad \frac{\D l^\mu}{\D \tau} = a\,u^\mu 
\end{equation}

For a given point $x^\mu$, the Fermi-Walker \cite{Synge} coordinates with space origin at the given worldline are
\begin{description}
\item[The time] $\tau(x^\nu)$ is given as an implicit function by
  \begin{equation}  \label{E3}
  \left[x^\mu - z^\mu(\tau)\right]\,u_\mu(\tau) = 0
\end{equation}
\item[The space coordinate] is defined by
\begin{equation}  \label{E4}
 \xi = \left[x^\mu - z^\mu(\tau)\right]\,l_\mu(\tau)
\end{equation}
\end{description}

By differentiating these two equations, it easily follows that
\begin{equation}  \label{E5}
 u_\mu\,\D x^\mu = - \left[1 + a(\tau)\xi\right]\, \D\tau \,, \qquad \qquad  l_\mu\,\D x^\mu = \D\xi
\end{equation}
and, as $u^\mu$ and $l^\mu$ form an orthonormal base, $\quad \D x^\mu = u^\mu\,\left[1 + a(\tau)\xi\right]\, \D\tau + l_\mu \D\xi \,$ and the invariant interval is
\begin{equation}  \label{E6}
\D s^2 = \D\xi^2 - \left[1 + a(\tau)\xi\right]^2 \D\tau^2
\end{equation}
Notice that, after the transformation of coordinates $dT = a(\tau)\,$, $dX =  \xi + 1/a(\tau)\,$,  
it becomes Rindler's interval. 

\subsection{ Uniqueness \label{S1.2}}
The form of the invariant interval (\ref{E6}) is unique in 1+1 Minkowski spacetime provided that the reference frame is both synchronous and rigid.

Indeed, let $(x,t)$ be coordinates for such a reference frame. The invariant infinitesimal interval is then
$$ \D s^2 = g_{11}(x)\,\D x^2 + g_{22}(x,t)\,\D t^2 \,, \qquad {\rm with} \qquad g_{11}> 0\,, \quad g_{22}<0 $$
and, introducing a new space coordinate by $\D \xi = \sqrt{g_{11}(x)}\,\D x\,$, we easily obtain
\begin{equation}  \label{E8}
\D s^2 = \D \xi^2- e^{2\phi} \,\D t^2 \,, \qquad {\rm with} \qquad \phi = \phi(x,t) 
\end{equation}

Now all Christoffel symbols vanish except
$$ \Gamma^1_{22} = e^{2\phi} \,\phi_\xi\,, \qquad  \Gamma^2_{12} = \phi_\xi\,, \qquad  \Gamma^2_{22} = \phi_t\,, $$ 
where $\phi_\mu = \partial_\mu\phi\,$.

The flatness condition in 1+1 dimensions amounts to $R^1_{\;212} = 0\,$, that is
$$ e^{2\phi}\,\left(\phi_{\xi\xi} + \phi_\xi^2 \right) \equiv e^\phi\,\partial_\xi^2 e^\phi = 0  $$
and two functions $f(t)$ and $h(t)$ exist such that $\,e^\phi = h(t) + \xi\,f(t)$.

If $h(t)\neq 0$, define $\D\tau = h(t)\,\D t\,$ and $\,a(\tau) = f(t)/h(t)\,$, and the invariant interval becomes (\ref{E6}); whereas, if $h(t)= 0$, as the metric is non-degenerate, necessarily $\,f(t) \neq 0\,$ and, introducing $\D\tau = f(t)\,\D t\,$ the invariant interval becomes 
$$  \D s^2 = \D \xi^2- \xi^2 \,\D \tau^2 \,, $$
namely Rindler interval, which is a limit case of (\ref{E6}) for $a \rightarrow \infty\,$.

\subsection{The Fermi-Walker reference frame associated to $z^\mu(\tau)$ \label{S1.1}  }
The timelike worldline $\,\xi = $constant and $\tau \in\mathbb{R}\,$ represents the history of the place $\xi$ in the space of the reference  frame associated to the Fermi-Walker coordinates. In Lorentzian coordinates this worldline is
\begin{equation}  \label{E6a}
\varphi^\mu(\tau; \xi) = z^\mu(\tau) + \xi\,l^\mu(\tau) \,, \qquad \qquad \tau \in\mathbb{R}  
\end{equation}

\begin{itemize}
\item $\xi=0$ is the space point whose worldline is $z^\mu(\tau)\,$ i. e. the space origin.
\item The proper time rate at the place $\xi$ is 
$$  \D\hat\tau =  \left[1 + a(\tau)\xi\right]\,\D\tau $$
$\hat\tau$ is the time ticked by a stationary atomic clock at $\xi$ and it only coincides with $\tau$ at the origin. 

Furthermore, the readings of proper time $\hat\tau$ by two stationary clocks at two different places $\xi_1 \neq \xi_2\,$ do not keep synchronized. We shall thus use the {\em synchronous time} $\tau\,$ instead of the local proper time $\hat\tau$.
\item According to the infinitesimal invariant interval (\ref{E6}), the infinitesimal radar distance \cite{Born},\cite{Landau} is $ \,dl^2 = d\xi^2\,$, therefore the distance between the space points of coordinates $\xi_1$ and $\xi_2$ is $\,L =\left|\xi_1 -\xi_2\right|\,$. 
\item The factor $\,1 + a(\tau)\xi\,$ is relevant in connexion with the domain of the Fermi-Walker coordinates, which does not embrace the whole Minkowski spacetime. Indeed, the procedure to obtain the coordinates of a point relies on solving the implicit function (\ref{E1}), which requires that the $\tau$-derivative of the left hand side does not vanish, that is $\,1 + a(\tau)\xi\neq 0 \,$.
\item The unit velocity vector (with respect to a Lorentzian frame) of the worldline $\xi =\,$constant in the Fermi-Walker reference frame, 
at the synchronous time $\tau$ is
$$ \frac{1}{\left|1 + a(\tau)\xi\right|}\, \partial_\tau \varphi^\mu(\tau;\xi) = u^\mu(\tau) \, \sign[1 + a(\tau)\xi] $$
and, to avoid time reversal, we shall restrict the domain of the Fermi-Walker coordinates to the region $\,1 + a(\tau)\xi > 0\,$, that is
$$ \mathcal{D} = \left\{ x^\mu = z^\mu(\tau) +\xi\, l^\mu(\tau)\,, \quad \xi > -1/a(\tau)\,, \quad \tau\in\mathbb{R}  \right\}  $$

The curve $\,1 + a(\tau)\xi = 0\,$ or, in Lorentzian coordinates $\,1 + a_\mu(\tau)\,\left[x^\mu - z^\mu(\tau)\right] = 0\,$ is the horizon of the Fermi-Walker frame. Specifically, in the case $a(\tau)= $constant and positive, $\mathcal{D}$ is the shadowed region in the figure below.
\begin{figure}[htbf]   \label{F1}
\begin{center} 
\includegraphics[height=5cm]{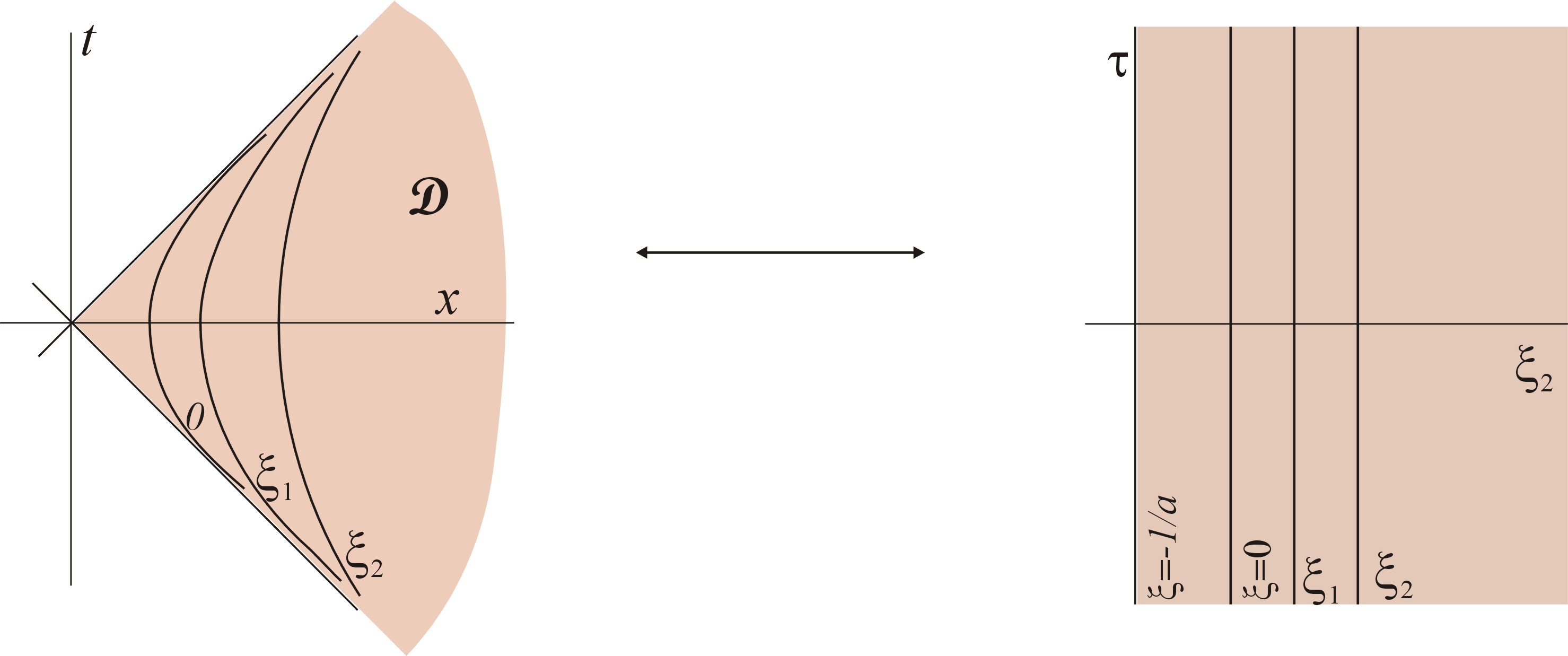}
\end{center}
\end{figure}

In the case $a(\tau)= $constant, this scheme recalls what is known as {\em Rindler space} \cite{Rindler}. Indeed, after the coordinate transformation 
$$\,X = \xi + \frac1{a} \,, \qquad \qquad T = a\,\tau $$
the invariant interval (\ref{E6}) becomes $\,\D s^2 = \D X^2 -X^2\,\D T^2 $. 

\item As for the proper acceleration of the worldline at $\xi$, we have
\begin{equation}  \label{E7}
a^\mu(\tau;\xi) = \frac{\D u^\mu}{\D\tau}\,\frac{\D\tau}{\D\hat\tau} = \frac{1}{1+\xi \,a(\tau)}\, a^\mu(\tau)
\end{equation}
and the invariant proper acceleration is
\begin{equation}   \label{E6aa}
a(\tau;\xi) = \frac{a(\tau)}{1+\xi \,a(\tau)} 
\end{equation}
which is different from place to place. As the latter can be solved for    
$$ \xi = \frac1{a(\tau;\xi)} - \frac1{a(\tau)} \, $$
we have that the space coordinate $\xi$ can be determined either by geometrical means with standard rods or by means of accelerometers, substracting the inverse proper accelerations of the place $\xi$ and the origin.
\end{itemize}

It is worth to mention here that Einstein's statement \cite{Einstein1913b}  
\begin{quotation}
 `` ... {\em acceleration} possesses as little absolute physical meaning as {\em velocity} '' 
\end{quotation}
does not hold {\em avant la lettre}. As a matter of fact, every place $\xi$ in the space of a Fermi-Walker reference frame has a proper acceleration (\ref{E6aa}) which is measurable with an accelerometer. However, the laws of classical particle dynamics also hold in the accelerated reference frame provided that a field of {\em inertial} force $-a^\mu(\tau;\xi)$ is included; the passive charge for this field being the inertial mass of the particle. It is only with this specification that two accelerated reference frames are equivalent from the dynamical (or even physical) viewpoint.

Also notice that, as local proper acceleration is different from place to place, there is not such a thing as {\em the acceleration} of a Fermi-Walker reference frame. However we shall use this expression to refer to the origin acceleration. 

\section{Uniformly accelerated reference frames \label{S2}}
For the sake of simplicity and to deal with closed expressions and elementary functions, we shall now restrict to the constant acceleration case $\,a(\tau) = a\,$. 

Assuming that the origin of the Fermi-Walker frame is initially at rest and coincides with the origin of the Lorentzian frame at $\tau =0\,$, we have that the rapidity is $\,\zeta = a \tau\,$ and the velocity vector and the origin worldline are respectively
\begin{equation} \label{E9}
u^\mu(\tau) = (\sinh \,a\tau, \cosh a\tau ) \,, \qquad    z^\mu(\tau) = \left(\frac1a\,\left[\cosh a\tau - 1 \right], \frac1a\,\sinh a\tau \right) \,, 
\end{equation}
Then, using equation (\ref{E6a}), the transformation formulae relating the Lorentzian coordinates $(x,t)$ and the Fermi-Walker coordinates $(\xi,\tau)$ of an event are
\begin{equation} \label{E10a}
t =\left(\xi + \frac1a \right)\,\sinh\,a\tau \,, \qquad \qquad x = - \frac1a + \left(\xi + \frac1a \right)\,\cosh\,a\tau
\end{equation}
and conversely
\begin{equation} \label{E10b}
\xi =  - \frac1a \pm \sqrt{ \left(x + \frac1a \right)^2 - t^2 } \,, \qquad \qquad \tau = \frac1a\, \artanh \frac{t}{x + \frac1a}
\end{equation}

\subsection{The class of uniformly accelerated reference frames \label{S2.1} }
Consider two uniformly accelerated reference frames, respectively $\mathcal{K}$ 
and $\mathcal{K}^\prime$,
the accelerations of the respective origins being $\,a\,$ and $\,a^\prime\,$, and let us find the transformation formulae relating the respective coordinates 
$$ (\xi,\tau) \quad \longrightarrow  \quad (\xi^\prime,\tau^\prime) $$
For elementary reasons it is obvious that these formulae must exhibit the group property.

It will exist a Lorentzian reference frame $\mathcal{L}$, with coordinates $(x,t)$, such that the origins of $\mathcal{L}$  and $\mathcal{K}$ coincide at rest when $t=\tau =0$. Similarly for $\mathcal{K}^\prime$ and a Lorentzian frame $\mathcal{L}^\prime$, with coordinates $(x^\prime,t^\prime)\,$.

The coordinates $(x,t)$ and $(\xi,\tau)$ are connected by equations (\ref{E10a}), the inverse transformation being given by (\ref{E10b}). Similar formulae, with ``primes'' added in the convenient places, hold for the coordinates $(x^\prime,t^\prime)$ and $(\xi^\prime,\tau^\prime)$.

On their turn, the reference frames $\mathcal{L}$ and $\mathcal{L}^\prime$ being Lorentzian, they are connected by a Poincaré transformation:
\begin{equation} \label{E11a}
x^\prime = (x-x_0)\,\cosh \kappa - (t-t_0)\,\sinh \kappa \,, \qquad \qquad 
t^\prime = - (x-x_0)\,\sinh \kappa + (t-t_0)\,\cosh \kappa \,, 
\end{equation}
and conversely
\begin{equation} \label{E11b}
x = x_0+ x^\prime\,\cosh \kappa +t^\prime\,\sinh \kappa \,, \qquad \qquad 
t = t_0 + x^\prime\,\sinh \kappa  + t^\prime\,\cosh \kappa \,, 
\end{equation}

Now, by successive application of the transformations (\ref{E10a}), (\ref{E11a}) and the primed equation (\ref{E10b}) (precisely in this order)
\begin{equation} \label{E11c}
 (\xi,\tau) \quad\stackrel{1}{\longrightarrow} \quad (x,t) \quad \stackrel{2}{\longrightarrow} \quad (x^\prime,t^\prime ) \quad \stackrel{3}{\longrightarrow} \quad (\xi^\prime,\tau^\prime) \,,
\end{equation}
we obtain the general transformation connecting the coordinates of two uniformly accelerated frames, which after a little algebra turns out to be:
\begin{eqnarray}  
\xi^\prime &=& -\frac1{a^\prime} + \sigma_{a^\prime} \left\{ \left(\xi +\frac1a \right)^2 + 2\left(\xi +\frac1a \right)\left(\cosh\,a\tau\left[\frac{\cosh\kappa}{a^\prime} -\frac1a -x_0\right] + \sinh a\tau\,\left[t_0 -\frac{\cosh\kappa}{a^\prime}\right]\right) \right. \nonumber \\[3ex]   \label{E12a}
 & & + \left. \left(x_0 +\frac1a\right)^2 - t_0^2 + \frac{1}{a^{\prime 2}} - \frac{2}{a^\prime}\left(x_0 +\frac1a\right) \cosh\kappa + 
 \frac{2 t_0}{a^\prime}\,\sinh \kappa \right\}^{1/2} \\[3ex]    \label{E12b}
 \tau^\prime & = & \frac1{a^\prime}\, \artanh \left\{\frac{\left[x_0 + \frac1a-\left(\xi+\frac1a \right)\cosh a\tau \right] \sinh\kappa - \left[t_0 - 
\left(\xi+\frac1a \right)\sinh a\tau \right] \cosh \kappa}{\frac1{a^\prime} - \left[x_0 + \frac1a-\left(\xi+\frac1a \right)\cosh a\tau \right] \cosh\kappa + \left[t_0 - \left(\xi+\frac1a \right)\sinh a\tau \right] \sinh \kappa  }\right\}
\end{eqnarray}
with $\sigma_{a^\prime} = \sign a^\prime\,$.

A short inspection reveals that, albeit the Poincaré group parameters, $x_0$, $t_0$ and $\kappa$, occur as the parameters of a group of transformations usually do, the two parameters $a$ and $a^\prime$ do not and they cannot be condensed in only one parameter (e. g. relative acceleration). Hence, in order that these expressions represent the action of a group of transformations on a manifold, the coordinates of the latter should include the  acceleration  $a$ of the origin besides the spacetime coordinates $\xi$ and $\tau$. The domain should rather be
$$  \mathcal{D} = \left\{ (\xi,\,\tau,\,a) \in \mathbb{R}^3\, ; \, 1 + a \,\xi > 0 \right\} $$
and the scheme (\ref{E11c}) should read 
$$ (\xi,\tau,a) \quad\stackrel{\psi}{\longrightarrow} \quad (x,t,0) \quad \stackrel{L}{\longrightarrow} \quad (x^\prime,t^\prime,0 ) \quad \stackrel{\psi^{\prime \,-1}}{\longrightarrow} \quad (\xi^\prime,\tau^\prime,a^\prime) \,,$$
If we then introduce
\begin{equation}   \label{E13a}
\rho = \frac1{a^\prime} - \frac1a \,, \qquad \qquad {\rm or} \qquad \qquad  a^\prime =\frac{a}{1 + a\rho} \,, 
\end{equation}
equations (\ref{E12a}-\ref{E12b}) become
\begin{eqnarray}  
\xi^\prime &=& - \frac{1+a\rho}{a} + \left(x_0 -\rho \right)^2 - t_0^2  + \sigma_{a^\prime} \left\{ \left(\xi +\frac1a \right)^2 + 2\left(\xi +\frac1a \right)\left(\cosh\,a\tau\left[\frac{(1+a\rho)\,\cosh\kappa}{a} -\frac1a -x_0\right]  \right. \right. \nonumber \\[3ex]   \label{E13b}
 & & + \left. \left .\sinh a\tau\,\left[t_0 -\frac{(1+a\rho)\,\cosh\kappa}{a}\right]\right) 
- \frac{2 (1+a\rho)}{a}\,\left[\left(x_0 +\frac1a\right) \cosh\kappa - t_0\,\sinh \kappa\right] \right\}^{1/2} \\[3ex]    \label{E13c}
 \tau^\prime & = & \frac{1+a\rho}{a}\, \artanh \left\{\frac{\left[x_0 + \frac1a-\left(\xi+\frac1a \right)\cosh a\tau \right] \sinh\kappa - \left[t_0 - \left(\xi+\frac1a \right)\sinh a\tau \right] \cosh \kappa}{\frac{1+a\rho}{a} - \left[x_0 + \frac1a-\left(\xi+\frac1a \right)\cosh a\tau \right] \cosh\kappa + \left[t_0 - \left(\xi+\frac1a \right)\sinh a\tau \right] \sinh \kappa  }\right\}
\end{eqnarray}

Now these three expressions, (\ref{E13a}) to (\ref{E13c}), have the shape of the action of a group on the manifold $\,\mathcal{D}\,$:
$$ y^{\prime i} = \Phi^i\left(y^j; \,p^A\right) \,, \qquad \qquad i,j = 1\ldots 3 \,, A = 1 \ldots 4 $$
where $y^i = (\xi,\tau,a)$ are the manifold coordinates, $p^A =(x_0,t_0,\kappa,\rho)$ are the group parameters. The identity transformation
corresponds to the parameters $x_0=t_0=\kappa=\rho=0$.

The infinitesimal generators for such a transformation law are 
$$ \mathbf{X}_A = \sum_j \left(\frac{\partial \Phi^j}{\partial p^A} \right)_{p^B = 0}\, \frac{\partial \;\;}{\partial y^j}  \quad ,$$ 
that is, vector fields on $\,\mathcal{D} $. A short calculation yields that the respective generators are
\begin{eqnarray}   \label{E15a}
\mathbf{P}_x & = & - \cosh a\tau\,\partial_\xi + \frac{\sinh a\tau}{1+a\xi}\,\partial_\tau \\[2ex] \label{E15b}
\mathbf{P}_t & = & \sinh a\tau\,\partial_\xi - \frac{\cosh a\tau}{1+a\xi}\,\partial_\tau      \\[2ex] \label{E15c}
\mathbf{J} & = & - \frac1a\,\mathbf{X}_t - \frac1a\,\partial_\tau    \\[2ex] \label{E15d}
\mathbf{D}_0 & = & - \mathbf{X}_x + a\tau\,\partial_\tau - a^2 \, \partial_a - \partial_\xi
\end{eqnarray}
and the commutation relations are:
\begin{equation}   \label{E16}
\left.  \begin{array}{ccc}
  \left[\mathbf{P}_x,\mathbf{P}_t\right] = 0 \qquad & \qquad \left[\mathbf{P}_x,\mathbf{J}\right] = - \mathbf{P}_t  \qquad &
	 \qquad \left[\mathbf{P}_t,\mathbf{J}\right] = - \mathbf{P}_x \\[2ex]
	\left[\mathbf{P}_x,\mathbf{D}_0\right] = 0 \qquad & \qquad \left[\mathbf{P}_t,\mathbf{D}_0\right] = 0  \qquad &
	 \qquad \left[\mathbf{J},\mathbf{D}_0\right] = 0 
        \end{array}   \right\}
\end{equation}
	
The Lie algebra generated by these vector fields is an abelian extension of 1+1 Poincaré algebra;
the direct sum of Poincaré algebra (generated by $\mathbf{P}_x\,$, $\mathbf{P}_t\,$ and $\mathbf{J}\,$), on the one hand, and $\,\mathbf{D}_0 \,$, on the other. 

This fact is consistent with the definition 
$$ \Phi = \psi^{\prime\,-1} \circ L \circ \psi $$
because, as it is clearly illustrated by the following diagram\\
\begin{center}
\includegraphics[height=5cm]{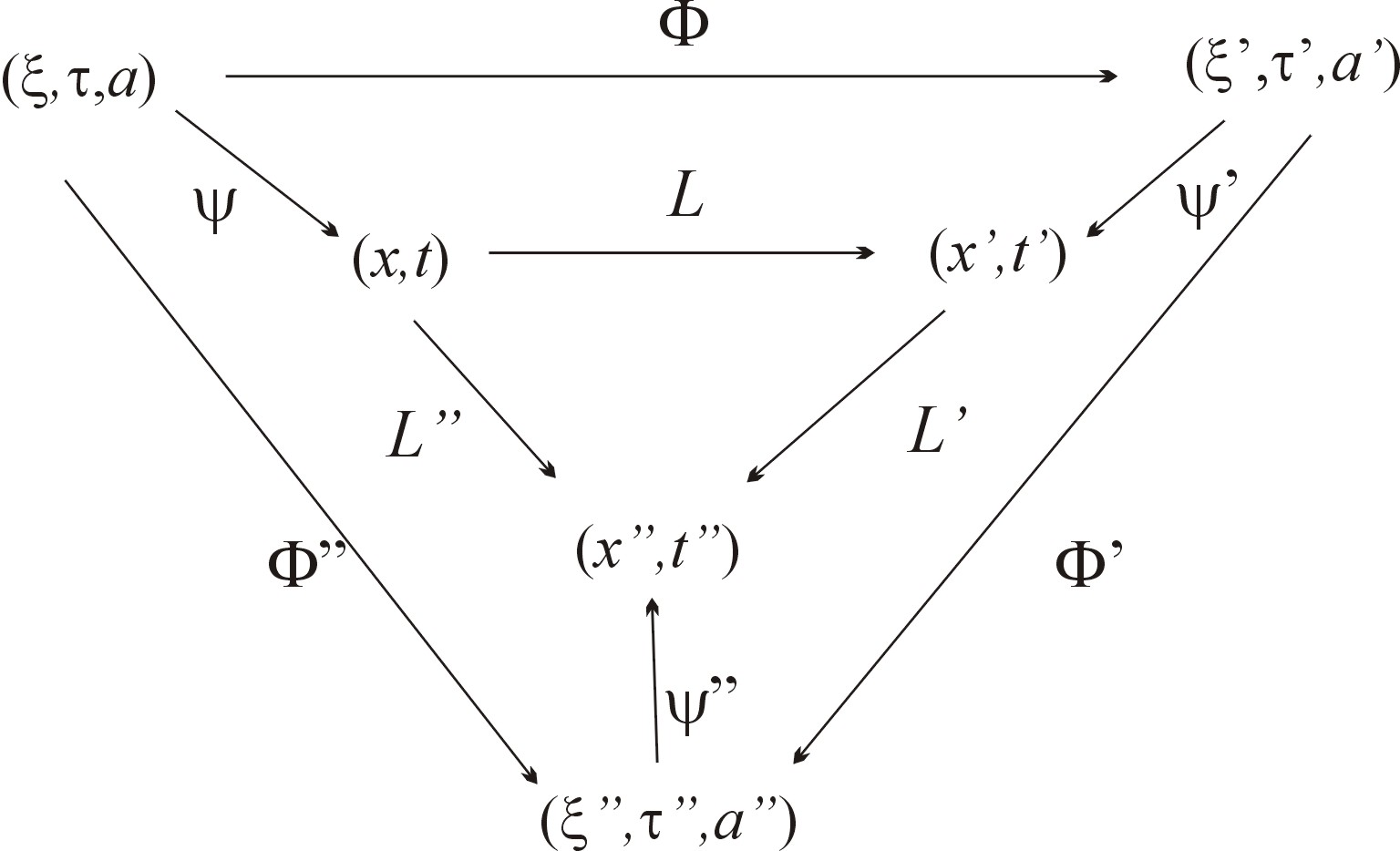}
\end{center} 
it follows that also
$$ \Phi^\prime = \psi^{\prime \prime\,-1} \circ L^\prime \circ \psi^\prime \qquad {\rm and} \qquad   \Phi^{\prime \prime} = \Phi^\prime \circ \Phi = \psi^{\prime \prime\,-1} \circ L^\prime \circ L \circ \psi    $$
and therefore $\,L^{\prime \prime} = L^\prime \circ L \,$, which means that the Poincaré group parameters $x_0,\,t_0,\, \kappa\,$ do not couple with the acceleration parameter $\rho\,$.

\section{Generalized isometries \label{S.3} }
In the preceeding section we have restricted to the case $a=$constant, in order to handle expressions in closed form and to be able to derive an explicit law of transformation from which the infinitesimal generators have been derived. The case $a(\tau)$  variable and arbitrary is more complex by far and will require some extra wit. To this end we shall exploit the notion of {\em generalized isometry} advanced by Bel in \cite{Bel93}.

In any accelerated frame the invariant interval has the generic form (\ref{E6})  
\begin{equation}  \label{E6r}
\D s^2 = \D\xi^2 - \left[1 + a(\tau)\xi\right]^2 \D\tau^2
\end{equation}
where $a(\tau)$ is some arbitrary function of one variable. Therefore, the transformation formulae $\,(\xi,\tau) \longleftrightarrow (\xi^\prime,\tau^\prime)\, $ connecting any two accelerated frames must preserve the form (\ref{E6r}), perhaps with two different functions $a(\tau) $ and $a^\prime(\tau^\prime)\,$, i. e. it si not an isometry because the metric is not invariant but {\em almost invariant} because only the function $a(\tau)$ changes whereas the form is preserved. 

To find the infinitesimal generalized isometries ---or generalized Killing vectors--- we write the interval as 
$$ \D s^2 = g_{\alpha\beta}(x, a_B(x)) \D x^\alpha\,\D x^\beta \,, \qquad $$
$a_B(x)\,,\quad  B = 1 \ldots m\,$ being a number of arbitrary functions ---in our particular case $m=1$--- and consider the infinitesimal transformations
$$ x^{\prime \alpha} = x^\alpha + \varepsilon\,X^\alpha(x) \,, \qquad a^\prime_B(x) = a_B(x) + \varepsilon\,A_B(x) $$
Then, as 
$$ g_{\mu\nu}(x^\prime, a^\prime_B(x^\prime)) \D x^{\prime \mu}\,\D x^{\prime \nu} = g_{\alpha\beta}(x, a_B(x)) \D x^\alpha\,\D x^\beta $$
and keeping only first order terms in $\,\varepsilon\,$, we have that
$$ X^\alpha \partial_\alpha g_{\mu \nu}\left( x, a_B(x) \right) + 2\,  X^\alpha_{\;|(\nu} \, g_{\mu)\alpha} \left( x, a_B(x) \right) + \sum_B A_B \left(\frac{\partial g_{\mu\nu}}{\partial a_B}\right)_{\left( x, a_B(x) \right)} = 0  $$
where ``$|$'' means partial derivative. Or, equivalently,
\begin{equation}  \label{E17}
 X_{\mu\|\nu} + X_{\nu\|\mu} + \sum_B A_B \frac{\partial g_{\mu\nu}}{\partial a_B} = 0
\end{equation}
which is the generalized Killing equation \cite{Bel93}.

In the particular case of the interval (\ref{E6}), which contains only one function $a(\tau)$ depending on only one variable,  we have that the above equation reduces to
$$ X^\alpha \partial_\alpha g_{\mu \nu}\left( x, a(\tau) \right) + 2 \,  X^\alpha_{\;|(\nu} \, g_{\mu)\alpha} + A(\tau)\,\partial_a g_{\mu \nu}\left( x, a(\tau) \right) = 0 \,,$$
that is
\begin{eqnarray}   \label{E18a}
\mu=\nu=1 \hspace*{1em} &\,, \quad\qquad  & \partial_\xi X^\xi = 0 \\[1ex]  \label{E18b}
\mu=1\,, \nu=2 &\,, \quad\qquad & \partial_\tau X^\xi - (1 + a \xi)^2\,   \partial_\xi X^\tau  = 0 \\[2ex]  \label{E18c}
\mu=\nu=2 \hspace*{1em} &\,,  \quad\qquad & \partial_\tau \left[(1 + a \xi)\,X^\tau\right] + a\,X^\xi + A\,\xi = 0 
\end{eqnarray}

The solution to the first equation is 
\begin{equation}   \label{E19}
X^\xi =  f(\tau) \,.
\end{equation}
where $f(\tau)$ is an arbitrary function. 
Including this, equation (\ref{E18b}) reads 
\begin{equation}   \label{E19a}
 \left[1+ \xi a(\tau)\right]^{2} \partial_\xi X^\tau = \dot f(\tau)  \,, 
\end{equation}
whose solution is
\begin{equation}  \label{E20}
X^\tau = - \frac{\dot f}{a ( 1 + \xi a)} + g 
\end{equation}
where $g=g(\tau)$ is an arbitrary function. 

Introducing then the solutions (\ref{E19}) and (\ref{E20}) into equation (\ref{E18c}) and including the separation of the variables $\xi$ and $\tau$, we readily obtain that
\begin{equation}  \label{E21a}
{\rm (a)}   \qquad  a\,f + \partial_\tau\left(g - \frac{\dot f}{a} \right) = 0  \,, \qquad \qquad 
 {\rm (b)}\qquad A + \partial_\tau \left( g \, a\right) = 0        
\end{equation}

If we now introduce the rapidity variable $\,\displaystyle{\zeta = \int_0^\tau \D t\,a(t)\,}$, equation (\ref{E21a}.a) becomes 
$$  \frac{\D^2 f}{\D\zeta^2} - f =  \frac{\D g}{\D\zeta} $$
whose general solution is
$$ f(\zeta) =\tilde{M}_+ e^\zeta + \tilde{M}_- e^{-\zeta} -g_0\,\sinh \zeta + \int_0^\zeta d\zeta^\prime\,g(\zeta^\prime)\,\cosh(\zeta-\zeta^\prime) \,, $$
where an integration by parts has been performed and $\tilde{M}_\pm$ are constant.

On its turn, equation (\ref{E21a}.b) becomes 
$$ \frac{\D \left(g\,a\right)}{\D\zeta} = - \frac{A}{a}  $$
which yields:
\begin{equation}  \label{E22a}
g(\zeta)= \frac{g_0}{a(\tau)} - \frac1{a(\tau)}\,\int_0^\tau \D t\,A(t) \,,  \qquad \qquad \zeta = \zeta(\tau) \,, \qquad g_0 = g(0)\, a(0)
\end{equation}
Substituting this in the above expression for $f(\zeta)$, after a little algebra we arrive at
\begin{eqnarray}     
f(\tau) &=& M_+ e^\zeta +  M_- e^{-\zeta} + \frac12\,g_0\,\left[e^{\zeta(\tau)} H^+(\tau) + e^{-\zeta(\tau)} H^-(\tau)   \right]
\nonumber \\[2ex]   \label{E22b}
 & & - \int_0^\tau \D t\,A(t) \,\frac12\,\left( e^{\zeta(\tau)} \left[H^+(\tau) - H^+(t) \right] + e^{-\zeta(\tau)} \left[H^-(\tau) - H^-(t) \right]  \right)
\end{eqnarray}
where the new constants $\displaystyle{M_\pm = \tilde{M}_\pm \mp \frac{g(0)}{2}   }\,$ have been introduced, and
\begin{equation}  \label{E22d}
 H^\pm(t) = \int_0^t \D t^\prime\,e^{\mp \zeta(t^\prime)} 
\end{equation}

Introducing now equations (\ref{E22a}) and (\ref{E22b}) into equations (\ref{E19}) and (\ref{E20}), we obtain that the infinitesimal generator is
\begin{eqnarray}
\mathbf{X} & = & f(\tau)\,\partial_\xi + \left[g(\tau)- \frac{\dot f(\tau)}{a(1+a\xi)}\right] \,\partial_\tau + \int_\mathbb{R} \D t \,A(t)\, \frac{\delta\quad}{\delta a(t)} \nonumber \\[3ex]  \label{E22e}
  & = & M_+\,\mathbf{P}_+ + M_-\,\mathbf{P}_- + g(0)\,\mathbf{J} +  \int_\mathbb{R} \D t \,A(t)\, \mathbf{D}_t
\end{eqnarray}
This generator acts on a manifold coordinated by $\,(\xi,\tau,[a(t)])\,$, where $\,a \in \mathcal{C}^0(\mathbb{R})\,$ and $\,(\xi,\tau) \in \mathbb{R}^2\,$ such that $\,1+a(\tau)\,\xi > 0 \,$.

Including now equations (\ref{E22a}-\ref{E22d}), we readily obtain that
\begin{eqnarray}  \label{E23a}   
\mathbf{P}_\pm  &=& e^{\pm \zeta(\tau)}\,\left[\partial_\xi \mp \frac1{1+\xi \,a(\tau)}\,\partial_\tau  \right]  \\[2ex]  \label{E23b}   
\mathbf{J}  &=& \frac{\xi}{1+\xi a(\tau)}\,\partial_\tau + \frac12\,H^+(\tau)\,\mathbf{P}_+ + \frac12\,H^-(\tau)\,\mathbf{P}_- 
\\[3ex]  \label{E23c} 
\mathbf{D}_t  &=& \frac{\delta\quad}{\delta a(t)} - \chi(t,\tau)\, \mathbf{J} + \chi(t,\tau)\,\frac12\,\left[H^+(t)\,\mathbf{P}_+ + H^-(t)\,\mathbf{P}_-\right]
\end{eqnarray}
with $\quad \chi(t,\tau)= \theta(t)\theta(\tau-t) - \theta(-t)\theta(t-\tau)\,$. 

Finally a straightforward albeit lengthy calculation permits to obtain the commutation relations
\begin{equation}  \label{E24}
\left.
\begin{array}{lll}
\left[\mathbf{P}_+,\mathbf{P}_- \right] = 0\,, \qquad & \qquad \left[\mathbf{J},\mathbf{P}_\pm  \right] = \pm \mathbf{P}_\pm 
\,,\qquad &   \\[2ex]
\left[\mathbf{D}_t,\mathbf{P}_\pm  \right] = 0 \,, \qquad & \qquad   \left[\mathbf{D}_t,\mathbf{J} \right] = 0   \,, \qquad & \qquad  
 \left[\mathbf{D}_{t},\mathbf{D}_{t^\prime}  \right] = 0  
\end{array}   \right\}
\end{equation}

Notice that the generators (\ref{E23c}) are parametrized by the continuous parameter $t\in \mathbb{R}$. This, together with the above commutation relations, means that the class of all generators (\ref{E22e}), that is the solution of the generalised Killing equation, is an infinite dimensional abelian extension of the Poincaré algebra in 1+1 dimensions.

As some authors prefer a discrete set of indices rather than a continuous parametrization of generators, in the Appendix we give an alternative, discrete set of generators.

To compare the above generators (\ref{E23a}-\ref{E23c}) with those obtained in the case of uniformly accelerated systems (Section \ref{S2}), it is convenient to redefine them as
$$ \mathbf{P}_x = -\frac12\,\left( \mathbf{P}_+ + \mathbf{P}_-\right) \,, \qquad \qquad 
\mathbf{P}_t = \frac12\,\left( \mathbf{P}_+ - \mathbf{P}_-\right)  $$
and then we have that:
\begin{eqnarray}  \label{E24a}   
\mathbf{P}_x  &=& -\cosh \zeta(\tau)\,\partial_\xi + \frac{\sinh \zeta(\tau)}{1+\xi \,a(\tau)}\,\partial_\tau    \,, \qquad
 \mathbf{P}_t  = \sinh \zeta(\tau)\,\partial_\xi - \frac{\cosh \zeta(\tau)}{1+\xi \,a(\tau)} \,\partial_\tau      \\[2ex]    \label{E24b}
\mathbf{J}  &=& C(\tau) \, \partial_\xi + \frac1{1+\xi a(\tau)}\,\left[ \xi - S(\tau)\right]\,  \partial_\tau     \\[3ex]  \label{E24c} 
\mathbf{D}_t  &=& \frac{\delta\quad}{\delta a(t)} + \chi(t,\tau)\, \left\{ \left[C(t)-C(\tau)\right]\, \partial_\xi - \frac1{1+\xi a(\tau)}\,\left[ \xi - S(\tau) + S(t) \right]\,  \partial_\tau \, \right)
\end{eqnarray}
where:
$$  C(t) = \int_0^t \D t^\prime\, \cosh\left[\zeta(\tau) - \zeta(t^\prime)\right] \,, \qquad \qquad 
 S(t) = \int_0^t \D t^\prime\, \sinh\left[\zeta(\tau) - \zeta(t^\prime)\right]  $$

If we now substitute $\zeta(\tau) = a \tau$, it follows immediately that (\ref{E24a}-\ref{E24b}) yield the generators (\ref{E15a}-\ref{E15c}), whereas the generator $\mathbf{D}_0$ in (\ref{E15d}) corresponds to $n=0$ in the discrete set of generators (\ref{A3}) (see the Appendix).  

\section{Conclusion}  
On the basis of the Fermi-Walker coordinates associated to a given worldline, we have approached the definition of accelerated reference frames in a 1+1 spacetime. We have restricted our work to only one space dimension to avoid the obstructions associated with rotational motion in relativity). We have proved the uniqueness of such a class of reference frames, provided that they are rigid and synchronous. (However, our results also hold in 3+1 dimensions, provided that the accelerations keep all in the same direction.)

In the case of steady acceleration, we have derived the transformation formulae relating two of such frames in explicit form and then we have obtained the infinitesimal generators, which constitute an abelian extension of Poincaré algebra in 1+1 spacetime. However, unlike the null acceleration case, these transformations do not apply to the whole spacetime and some horizons occur; the domains of validity usually depending on the magnitude of the acceleration.

The general case, namely arbitrary acceleration, does not allow to obtain closed, explicit expressions for the transformation formulae, but we have succeeded in studying the infinitesimal transformations for arbitrarily accelerated reference systems. To this end we have resorted to the notions of generalised isometries and generalised Killing equation;  infinitesimal transformations are then associated to the generalized Killing vectors of the spacetime metric (\ref{E6}). The latter constitute an infinite dimensional Lie algebra which turns out to be an abelian extension of Poincaré algebra that, as expected, also contains the above mentioned extension for steadily accelerated motions. 

As the infinite Lie algebra we have obtained is an abelian extension, our result is rather dull from a mathematical point of view. Things could turn more involved, hence more interesting, if the number of space dimensions is increased. So, in future work we shall exploit the techniques based on the notion of generalised isometry to explore the infinitesimal transformations connecting two reference systems with arbitrary accelerations in ordinary spacetime ---first for non-rotating systems and then we will introduce arbitrary rotational motion---, aiming to extend Poincaré algebra along he lines developed here.

\section*{Acknowledgement}
The author thanks A. Molina and L. Bel for their valuable suggestions and stimulating discussion.

\section*{Appendix}
To translate the continuously parametrized generators $\mathbf{D}_t$ (\ref{E23c}) into a discrete infinite set of generators, we shall replace $a(t)$ by its [formal] infinite Taylor polynomial $\displaystyle{a(t) = \sum_{n=0}^\infty \frac{a_n}{n!}\, t^n }\,$, thus defining a 1 to 1 correspondence $\,(a_n)_{n\in\mathbb{N}} \longrightarrow a(t) \,$, whose inverse is
$$ a_n = (-1)^n\,\int_\mathbb{R} \D t\,a(t)\,\delta^{(n)}(t)  $$
Therefore $\displaystyle{ \frac{\delta a_n}{\delta a(t)} =  (-1)^n\,\delta^{(n)}(t)  }$ and we have the [formal] relation
\begin{equation}   \label{A1}
\frac{\delta \;\;}{\delta a(t)} =  \sum_n (-1)^n\,\delta^{(n)}(t)\, \frac{\partial \; \;}{\partial a_n}   \qquad {\rm and} \qquad 
\frac{\partial \; \;}{\partial a_n} = \frac1{n!}\,\int_\mathbb{R} \D t\, t^n\,\frac{\delta \;\;}{\delta a(t)}
\end{equation}

Substituting now $A(t)$ for its formal Taylor expansion we have
\begin{equation}   \label{A2}
\int_\mathbb{R} \D t \,A(t)\, \mathbf{D}_t = \sum_n A_n\,\mathbf{D}_n \,, \qquad {\rm with} \qquad 
\mathbf{D}_n = \frac1{n!}\,\int_\mathbb{R} \D t\, t^n\, \mathbf{D}_t
\end{equation}
which, including (\ref{A1}) and (\ref{E23c}), yields
\begin{equation}   \label{A3}
\mathbf{D}_n = \frac{\partial \; \;}{\partial a_n} - \frac{\tau^{n+1}}{(n+1)!}\,\mathbf{J} + L^+_n(\tau) \mathbf{P}_+ +  
L^-_n(\tau) \mathbf{P}_- 
\end{equation}
where
$$ L^\pm_n(\tau) = \frac1{2(n+1)!}\,\int_0^\tau \D t \,e^{\mp \zeta(t)}\,\left(\tau^{n+1} - t^{n+1}\right)  $$
As for the commutation relations, it follows from (\ref{E24}) that
$$ \left[\mathbf{D}_n,\mathbf{P}_\pm  \right] =    \left[\mathbf{D}_n,\mathbf{J} \right] = 
 \left[\mathbf{D}_{n},\mathbf{D}_m  \right] = 0  \,, \qquad \forall n, \, m \in \mathbb{N}   $$

Notice that the subalgebra generated by $\mathbf{D}_0,\,\mathbf{P}_\pm, \, \mathbf{J}$ coincides with the constant acceleration extension of Poincaré algebra given by the relations (\ref{E16}).

\end{document}